\documentclass[12pt]{iopart}
\usepackage{epsfig}
\usepackage{iopams}
\usepackage{amsfonts} 
\newcommand{\seq}{\begin{subequations}} 
\newcommand{\sen}{\end{subequations}} 
\newcommand{\eq}{\begin{eqnarray}} 
\newcommand{\en}{\end{eqnarray}} 
\newcommand{\ra}{\rangle}

\def\dppm{D^\pm} 
\def\dpp{D^+} 
\def\dpm{D^-} 
\def\dpmps{D^{\ast \mp}} 
\def\dpps{D^{\ast +}} 
\def\dpms{D^{\ast -}}

\def\d{D^0} 
\def\db{\bar{D}^{0}} 
\def\ds{D^{\ast  0}} 
 
\def\dbs{\bar{D}^{\ast 0}}

\def\L2{\Lambda^2} 
 
\def\jp{J/\psi}  
\def\jpsi{J_\psi}

\begin{document} 
 
\title{$J/\psi\gamma$ and $\psi(2S)\gamma$ decay modes of the $X(3872)$}

\author{\hspace*{-3cm}
Yubing~Dong$^{1,2,3}$,~Amand~Faessler$^3$,~Thomas~Gutsche$^3$,~Valery~E~Lyubovitskij$^3$\footnote{On leave of absence 
        from Department of Physics, Tomsk State University, 
        634050 Tomsk, Russia} 
\vspace*{1.2\baselineskip}} 
\address{ 
$^1$ Institute of High Energy Physics, Beijing 100049, P. R. China\\  

$^2$ Theoretical Physics Center for Science Facilities (TPCSF), CAS,\\  
Beijing 100049, P. R. China \\ 

$^3$ Institut f\"ur Theoretische Physik,  Universit\"at T\"ubingen,\\ 
Kepler Center for Astro and Particle Physics, \\  
Auf der Morgenstelle 14, D-72076 T\"ubingen, Germany}  
 
\ead{
dongyb@ihep.ac.cn, 
amand.faessler@uni-tuebingen.de,\\
thomas.gutsche@uni-tuebingen.de,
valeri.lyubovitskij@uni-tuebingen.de} 
 
\begin{abstract}  
 
The $X(3872)$ with quantum numbers $J^{PC} = 1^{++}$ is considered 
as a composite state containing both molecular hadronic
and a $c\bar{c}$ component. Based on this structure assumption we 
first constrain model parameters in order to reproduce the 
predictions for the radiative decay widths of
$X(3872)\to J/\psi\gamma$ and $X(3872)\to \psi(2S)\gamma$ as obtained
in $c\bar c$ potential models.
Depending on these predictions we find
that further inclusion of the 
molecular component can in principle lead to an improved description
of the radiative $X(3872)$ decays.
We also show that strong decay modes of the $X(3872)$ and in particular
the ratio of radiative ($\jp\gamma$) to strong ($\jp\pi^+\pi^-$) decays
hint towards a subleading role of the $c\bar c$ component in the $X(3872)$.

\end{abstract} 

\vskip 1cm


 
\maketitle 
 
\newpage 
 
\section{Introduction} 
 
The $X(3872)$ is one of the peculiar new meson resonances which was discovered
during the last years~\cite{Nakamura:2010zz} and where its properties cannot  
be simply explained in the context of conventional constituent quark models.
Presently several structure interpretations for this new resonance are
proposed in the literature (for a status report see e.g. 
Refs.~\cite{Swanson:2006st,Bauer:2005yu,Voloshin:2007dx}). 
In the molecular approaches  
of~\cite{Voloshin:1976ap}-\cite{Fleming:2008yn} it is argued that the 
$X(3872)$ can be identified with a weakly--bound hadronic molecule 
whose constituents are $D$ and $D^\ast$ mesons. This natural interpretation 
is due to the fact that its mass $m_X$ is very close to the 
$\d \dbs$ threshold and hence is in analogy to the deuteron --- 
a weakly--bound state of proton and neutron.  
 
First it was proposed that the state $X(3872)$ is a superposition of  
$\d \dbs$ and $\db \ds$ pairs. Later (see e.g. discussions in 
Refs.~\cite{Swanson:2003tb,Voloshin:2004mh,Braaten:2005ai}) other
additional configurations
such as a charmonium or even other meson pair components
were discussed in addition to $\d \dbs +$ its charge conjugate (c.c.)
(Here and in the following we use the convention that $\dbs$ does not  
change sign under charge conjugation. For a detailed discussion see also
Ref.~\cite{Thomas:2008ja}). The additional possibility of two nearly degenerate
$X(3872)$ states with positive and negative charge parity has been 
discussed in Refs.~\cite{Gamermann:2007fi,Terasaki:2007uv}.  
 
In the present paper we focus on the radiative decays of the $X(3872)$. 
There exist already several calculations for the radiative decays of 
the $X(3872)$ and various approaches give quite different results for the 
partial decay widths even when based on the same structure assumption. 
A new measurement by the {\it BABAR} 
Collaboration gives clear evidence for a strong radiative decay mode
involving the $\psi(2S)$~\cite{Aubert:2008rn}. They indicate the measured
ratio of
\begin{eqnarray}\label{ratio} 
R = \frac{{\cal B}(X(3872)\rightarrow \psi(2S)\gamma)} 
{{\cal B}(X(3872)\rightarrow J/\psi\gamma)}=3.5\pm 1.4. 
\end{eqnarray}
A naive estimate for this ratio just involving phase space gives a suppression
of the $\psi(2S)\gamma$ relative to the $J/\psi\gamma$ decay mode.
Therefore, dynamical selection rules play an important role to explain
the enhanced $\psi(2S)\gamma$ mode.
A first model-dependent prediction for the radiative decays in the molecular
picture~\cite{Swanson:2004pp} 
indicates that while the decay to $J/\psi\gamma$ can be accommodated,
the width of the decay channel $\psi(2S)\gamma$ is very small.
It is therefore expected that the ratio of Eq.~(\ref{ratio})
gives some constraint on a possible charmonium component in the $X(3872)$.
Analyses of the radiative decays of the $X(3872)$ interpreted as a pure
$2^3P_1 $ charmonium configuration have been performed 
in potential approaches (see e.g. 
Refs.~\cite{Barnes:2003vb,Barnes:2005pb,Li:2009zu}). There it was shown 
that the results for the ratio $R$ vary from 1.3 to 12.8 depending in
particular for the $X \to \gamma J/\psi$ transition on
sensitive details such as the position of the node in 
the transition form factor.
Especially the result of Ref.~\cite{Barnes:2005pb} is compatible 
with current data including errors. But it should also be noted that potential
model predictions for the mass of the $2^3P_1 $ configuration are about
30~\cite{Li:2009zu} to 80 MeV~\cite{Barnes:2003vb} higher than the
observed $X(3872)$ mass.

In Refs.~\cite{Dong:2009gb,Dong:2008gb,Faessler:2007gv,Branz:2008cb}  
we developed a formalism for the study of recently observed  
exotic meson and baryon states 
(like $D_{s0}^\ast(2317)$, $D_{s1}(2460)$, $X(3872)$, 
$Y(3940)$, $Y(4140)$, $Z(4430)$, $\Lambda_c(2940)$,~$\Sigma_c(2800)$, 
$\ldots$) as hadronic molecules. In Ref.~\cite{Dong:2008gb}  
we extended our formalism to the description of the decay $X \to \jp \gamma$
assuming that the  
$X$ is a $S$--wave $(\d  \dbs + \ds \db)/\sqrt{2}$ molecule of
positive charge parity.
As in the case of the $D_{s0}^{\ast}$ and  $D_{s1}$ states
a composite (molecular) structure of the $X(3872)$ meson is defined  
by the compositeness condition $Z=0$~\cite{Weinberg:1962hj,% 
Efimov:1993ei,Anikin:1995cf} (see also 
Refs.~\cite{Dong:2009gb,Dong:2008gb,Faessler:2007gv,% 
Branz:2008cb,Ivanov:2005fd}). This condition  
implies that the renormalization constant of the hadron wave function  
is set equal to zero or that the hadron exists as a bound state of its  
constituents. The compositeness condition was originally  
applied to the study of the deuteron as a bound state of proton and 
neutron~\cite{Weinberg:1962hj}. Then it was extensively used 
in low--energy hadron phenomenology as the master equation for the 
treatment of mesons and baryons as bound states of light and heavy 
constituent quarks (see e.g. Refs.~\cite{Efimov:1993ei,Anikin:1995cf}).  
By constructing a phenomenological Lagrangian including the  
couplings of the bound state to its constituents and of the constituents  
to other particles in the possible decay channels we calculated meson--loop  
diagrams describing different decays of the molecular states  
(see details in~\cite{Dong:2009gb,Dong:2008gb,Faessler:2007gv,Branz:2008cb}).  
In Ref.~\cite{Dong:2008gb} we originally estimated the role of a possible 
charmonium component in the radiative $X(3872)$ decay. Our consideration 
was based on an effective hadronic Lagrangian describing the interaction of 
$J/\psi$ with $D$ and $D^\ast$ mesons taken from an analysis of $J/\psi$ 
absorption in hadronic matter~\cite{Lin:1999ad}. 
Then in Ref.~\cite{Dong:2009gb} we considered the $X(3872)$ as 
a superposition of the molecular $\d\ds$ 
component and other hadronic pairs -- $D^{\pm} D^{\ast\,\mp}$,  
$\jp\omega$ and $\jp\rho$. The strong couplings of the charmonium states 
$J/\psi$ and $\psi(2S)$ to $D$ and $D^*$ mesons were now consistently
taken from heavy hadron chiral perturbation theory 
(HHChPT)~\cite{Colangelo:2003sa}. 
In Ref.~\cite{Dong:2008gb} our formalism was 
based on the idea that the molecular $\d\ds$ component dominates the
radiative decay of the $X(3872)$. In this vein we used a dominant fraction 
of the $\d\ds$ component in the $X(3872)$ Fock state and also 
a relatively large value for the dimensional parameter $\Lambda_{DD^\ast}$ 
describing the distribution of $\d$ and $\ds$ mesons in the $X(3872)$. 
This led to the conclusion that the molecular $\d\ds$ component plays
the leading role when compared to an admixed charmonium configuration
in determining the $X \to \jp \gamma$ 
decay width~\cite{Dong:2009gb,Dong:2008gb}. 
Here we improve our approach by using a smaller value for the 
parameter $\Lambda_{DD^\ast} = 0.5$ GeV to generate a
$\d\ds$ molecular component which is spatially more extended in comparison
to the charmonium one.
Also, we modify the vertex functions of the involved excited charmonium
configurations in the $X(3872)$ and for the $\psi(2S)$ such as to match
the corresponding band width of predictions for radiative
transitions of heavy quarkonia in potential
models~\cite{Barnes:2003vb,Barnes:2005pb,Li:2009zu}. 
Both an inclusion of the molecular and the charmonium components 
in the $X(3872)$ can lead to an improved description of the ratio $R$ 
involving the radiative modes while still being
able to explain the strong decays of the $X(3872)$ such 
as $J/\psi \pi^+ \pi^-$ and $J/\psi \pi^+ \pi^- \pi^0$. 
Therefore, the main objective of our paper is to demonstrate 
that a more precise measurement of the radiative decays of $X(3872)$ 
and the corresponding ratio R can help to understand the composite 
structure of this unusual state, in particular to determine the fraction 
of the molecular and the charmonium components. 
 
In the present paper we proceed as follows. In Sec.~II we first discuss 
the basic notions of our approach. We discuss the effective Lagrangian 
for the treatment of the $X(3872)$ meson as a superposition of the 
molecular and charmonium components.  
Second, we consider the radiative two--body decays $X \to \psi(2S)\gamma$ and  
$X \to J/\psi\gamma$ decays. In Sec.~III we present our numerical results.  
In Sec.~IV we present a short summary. 
 
\section{Approach}  
 
Following Refs.~\cite{Swanson:2003tb,Dong:2009gb,Dong:2008gb} 
we consider the $X(3872)$ state as a superposition  
of the dominant molecular $\d\ds$ component, of other hadronic  
configurations--$D^{\pm}D^{*\mp}$, $J/\psi\omega$, $J/\psi\rho$, 
as well as of the $c\bar c$ charmonium configuration as 
\eq\label{Xstate} 
|X(3872)\ra &=& \cos\theta\Bigg [ 
\frac{Z_{\d\ds}^{1/2}}{\sqrt{2}}( |\d\dbs \ra + |\ds\db \ra )   
+ \frac{Z_{\dppm\dpmps}^{1/2}}{\sqrt{2}}( | \dpp\dpms \ra  
\nonumber \\ 
&+& | \dpm\dpps \ra )  + Z_{\jpsi\omega}^{1/2} | \jpsi \omega \ra  
 + Z_{\jpsi\rho}^{1/2} | \jpsi \rho \ra \Bigg ]
 + \sin\theta \, |c\bar{c}\ra \; .   
\en  
Here $\theta$ is the mixing angle between the hadronic and the charmonium 
components: $\cos^2\theta $ and $\sin^2\theta$ represent the probabilities  
to find a hadronic and charmonium configuration, respectively,
for the normalization  
\eq\label{Zfac}  
Z_{\d\ds} + Z_{\dppm\dpmps} + Z_{\jpsi\omega} +  Z_{\jpsi\rho} = 1 \,.  
\en   
The limiting case of $\cos\theta = 1$ corresponds to the situation
where $X(3872)$ has no charmonium component (this case was considered  
recently in Ref.~\cite{Dong:2009gb}) while the value $\cos\theta=0$
refers to the pure charmonium interpretation of the $X(3872)$.  
In this paper we will simply employ values for $Z_{\d\ds}$,  
$Z_{\dppm\dpmps}$, $Z_{\jpsi\omega}$ and $Z_{\jpsi\rho}$
as derived in a potential model~\cite {Swanson:2003tb}.  
 
Our approach is based on an effective interaction Lagrangian describing  
the couplings of the $X(3872)$ to its constituents. Following our 
formalism developed in~\cite{Dong:2009gb} we  
apply a nonlocal form containing the correlation functions $\Phi(y^2)$ 
characterizing the distribution of the constituents in the $X(3872)$.
The Lagrangian is set up as:
\eq\label{Lagr_X} 
{\cal L}_X(x) &=& g_{_{X\d\ds}} \, X_\mu(x) \, J^\mu_{\d\ds}(x)  
+ g_{_{X\dppm\dpmps}} \, X_\mu(x) \, J^\mu_{\dppm\dpmps}(x) 
\nonumber\\ 
&+& \frac{g_{_{X\jpsi\omega}}}{m_X} \,  
\epsilon_{\mu\nu\alpha\beta} \, X^\mu(x)  
\, J^{\beta\mu}_{\jpsi\omega}(x)  
+ \frac{g_{_{X\jpsi\rho}}}{m_X} \,  
\epsilon_{\mu\nu\alpha\beta} \, X^\mu(x)  
\, J^{\beta\mu}_{\jpsi\rho}(x)  \nonumber \\ 
&+&g_{_{Xc\bar{c}}}X_{\mu}(x)J^{\mu}_{c\bar{c}}(x) 
\,, 
\en  
where $g_{_{XH_1H_2}}$ and $g_{_{Xc\bar{c}}}$ are the couplings of $X(3872)$  
to the hadronic constituents $H_1 H_2$ and to the $c\bar{c}$ component.  
$X$ is the field describing $X(3872)$ and $J^\Gamma_{\psi_1\psi_2}$  
are the respective currents composed of the fields $\psi_1$ and $\psi_2$ with: 
\eq\label{current_nonlocal}  
\hspace*{-2.55cm}
J^\mu_{D\bar D^\ast}(x) &=& \frac{1}{\sqrt{2}}  
\int d^4y \Phi_{DD^\ast}(y^2) \biggl( D(x+y/2) \bar D^{\ast\mu}(x-y/2)  
+ \bar D(x+y/2) D^{\ast\mu}(x-y/2) \biggr) \,,\nonumber\\  
\hspace*{-2.55cm}
J^{\nu\alpha\beta}_{\jpsi V}(x)& = & \jpsi^\nu(x) \, 
\int d^4y \Phi_{V}(y^2) \partial^\alpha V^\beta(x+y) \,,\\ 
\hspace*{-2.55cm}
J^{\mu}_{c\bar{c}}(x)&=& 
\int d^4y \Phi_{c\bar{c}}(y^2)\bar{c}(x+y/2)\gamma^{\mu}\gamma^5c(x-y/2)\,. 
\nonumber 
\en  
Here $\Phi_{DD^\ast}$ is the correlation function describing the  
distribution of $D D^\ast$ inside $X$, the function $\Phi_{V}$ describes  
the distribution of the light vector meson $V=\rho$ or $\omega$ around  
the $J/\psi$, which is located at the center of mass of the $X(3872)$,  
and the function $\Phi_{c\bar{c}}$ represents the distribution of the  
$c\bar{c}$ component. A basic requirement for the choice of an explicit  
form of the correlation functions $\Phi_I$ $(I=DD^\ast,V,c\bar{c})$ is  that  
their Fourier transforms vanish sufficiently fast in the ultraviolet region  
of Euclidean space to render the Feynman diagrams ultraviolet finite. 
We adopt the Gaussian form for the correlation functions. In particular, 
for the molecular correlation functions the 
Fourier transform of $\Phi_I$ $(I = DD^\ast,V)$ reads:  
\eq  
\tilde\Phi_{I}(p_E^2/\Lambda^2_I) \doteq \exp( - p_E^2/\Lambda_I^2)\,, 
\en  
where $p_{E}$ is the Euclidean Jacobi momentum and $\Lambda_I$ 
is a size parameter. Here we use the values of 
$\Lambda_{DD^\ast} = 0.5$~GeV and $\Lambda_{V} \simeq m_V = 0.77$~GeV. 
For the correlation function of the charmonium component we use the 
modified Gaussian form multiplied by a polynomial:  
\eq  
\tilde\Phi_{c\bar c}(p_E^2/\Lambda^2_I) \doteq 
\exp( - p_E^2/\Lambda_{c\bar c}^2) (1 + \alpha_X p_E^2/\Lambda_{c\bar c}^2) 
\en 
where $\Lambda_{c\bar c} = 3.5$ GeV,  
$\alpha_X$ is a free parameter which will be fixed from the analysis 
of the radiative decays of $X(3872)$ in the pure charmonium case.

The coupling constants $g_{_{X\psi_1\psi_2}}$ are determined by the
compositeness condition~\cite{Weinberg:1962hj,Efimov:1993ei,Anikin:1995cf,%
Faessler:2007gv}. This requirement implies that the renormalization constant 
of the hadron wave function is set equal to zero with  
\eq\label{ZX} 
Z_X = 1 - \Sigma_X^\prime(m_X^2) = 0 \,. 
\en 
Here $\Sigma^\prime_X(m_{X}^2) = d\Sigma_X(p^2)/dp^2|_{p^2=m_X^2}$  
is the derivative of the transverse part of the mass operator  
$\Sigma^{\mu\nu}_X$ conventionally split into the transverse 
$\Sigma_X$ and longitudinal $\Sigma^L_X$  parts as: 
\eq 
\Sigma^{\mu\nu}_X(p) = g^{\mu\nu}_\perp \Sigma_X(p^2)  
+ \frac{p^\mu p^\nu}{p^2} \Sigma^L_X(p^2) \,, 
\en 
where  
$g^{\mu\nu}_\perp = g^{\mu\nu} - p^\mu p^\nu/p^2$  
and $g^{\mu\nu}_\perp p_\mu = 0\,.$  
The mass operator of the $X(3872)$ receives contributions
from five hadron--loop diagrams with
\eq  
\hspace*{-2cm}
\Sigma_X(m_X^2) = \Sigma_{\d\ds}(m_X^2) +  
\Sigma_{\dppm\dpmps}(m_X^2) + \Sigma_{\jpsi\omega}(m_X^2)  
+ \Sigma_{\jpsi\rho}(m_X^2) + \Sigma_{c\bar c}(m_X^2)\,, 
\en 
which are induced by the interaction of $X$ with the corresponding hadronic  
pairs $H_1H_2$ and the $c\bar c$ configuration as contained in 
Eq.~(\ref{Lagr_X}). The relevant diagram contributing to 
$\Sigma^{\mu\nu}_X(p)$ is shown in Fig.1.  
Using Eq.~(\ref{Xstate}) and the compositeness condition~(\ref{ZX})  
we have five independent equations to determine the  
coupling constants $g_{_{X\psi_1\psi_2}}$. The relative signs  
of the coupling constants are not fixed from the compositeness condition.
Details on the relevance of the signs are discussed in the section related
to the analysis of the radiative decays of the $X(3872)$.
Note, the $Z$ factors of Eq.~(\ref{Zfac}) cannot be directly identified with the couplings
of $X(3872)$ to the different channels but are related via the compositeness condition,
which is a standard result in field theoretical approaches. An attempt
to link the couplings to the concept of wave functions assuming zero range transition
operators has been performed in Ref.~\cite{Gamermann:2009uq} in a coupled channel formalism.
There it was shown that these couplings are essentially a measure of the wave function close
to the origin. 

In Fig.2 we display the diagrams relevant for the radiative decays
$X\to J/\psi\gamma$ and $X\to\psi(2S)\gamma$. The main  
difference between the sets of diagrams feeding the channels
$J/\psi\gamma$ and $\psi(2S)\gamma$ is: the last diagram of Fig.2 which 
is generated by the $J/\psi V$ $(V=\rho,\omega)$ component of  
the $X(3872)$ only contributes to the $J/\psi\gamma$ final state.
The effective Lagrangians generating the electromagnetic interaction 
vertices are  
\begin{eqnarray} 
\hspace*{-2cm}
{\cal L}_{DD\gamma}(x)&=&  
ie A_\mu(x) D^-(x) \!\stackrel{\leftrightarrow}{\partial}^{\,\mu} 
D^+(x) \,,   
\nonumber\\  
\hspace*{-2cm}
{\cal L}_{D^*D^*\gamma}(x)&=& - ie A_\mu(x) \, \biggl\{ g^{\alpha\beta} \,  
D^{*-}_\alpha \!\stackrel{\leftrightarrow}{\partial}^{\,\mu} 
 D^{*+}_\beta(x)  - g^{\mu\beta} \,  
D^{*-}_\alpha(x) \partial^\alpha D^{*+}_\beta(x)  \nonumber\\
&+& g^{\mu\alpha} \,  
\partial^\beta D^{*-}_\alpha(x) D^{*+}_\beta(x) \biggr\} \,, \nonumber\\  
\hspace*{-2cm}
{\cal L}_{D^*D\gamma}(x)&=& 
\frac{e}{4} \epsilon^{\mu\nu\alpha\beta}F_{\mu\nu}(x)  
\biggl\{  g_{D^{*-}D^+\gamma} D^{*-}_{\alpha\beta}(x)D^+(x)  
+ g_{D^{*0}D^0\gamma} \bar{D}^{*0}_{\alpha\beta}(x)D^0(x)  
\biggr\} +  {\rm H.c.} \,,\nonumber\\ 
\hspace*{-2cm}
{\cal L}_{cc\gamma}(x)&=&\frac{2e}{3}A_{\mu}(x)\bar{c}(x)\gamma^{\mu}c(x)\,,  
\end{eqnarray}   
where $A \!\stackrel{\leftrightarrow}{\partial}_{\mu} B 
= A \partial_{\mu} B  
- \partial_{\mu} A B$;  
$F_{\mu\nu}=\partial_{\mu}A_{\nu}-\partial_{\nu}A_{\mu}$, and  
$M_{\mu\nu}=\partial_{\mu}M_{\nu}-\partial_{\nu}M_{\mu}$.   
The strong couplings of the charmonium states $J/\psi$ and $\psi(2S)$ to  
$D$ and $D^*$ mesons are taken from HHChPT~\cite{Colangelo:2003sa} 
(see also Ref.~\cite{Dong:2009gb}).     
For convenience we use a relativistic normalization of the meson states  
and write the Lagrangians in manifestly Lorentz covariant form:  
\begin{eqnarray} 
\hspace*{-2cm}
{\cal L}_{\psi_n DD}(x)&=&-ig_{\psi_n DD} 
\psi_n^{\mu}D_i^{\dagger}(x)
\!\stackrel{\leftrightarrow}{\partial}_{\,\mu} D_i(x)\,, 
\nonumber \\ 
\hspace*{-2cm}
{\cal L}_{\psi_n DD^*}(x)&=&g_{\psi_n DD^*} \epsilon_{\mu\nu\alpha\beta}  
\partial^\mu\psi_n^{\nu} \biggl\{  
 D_i^{*\beta \dagger}(x) \!\stackrel{\leftrightarrow}{\partial}^{\alpha}D_i(x) 
-D_i^{\dagger}(x) 
\!\stackrel{\leftrightarrow}{\partial}^{\alpha}D^{*\beta}_i(x)  
\biggr\} \,,\\ 
\hspace*{-2cm}
{\cal L}_{\psi_n D^*D^*}(x)&=&ig_{\psi_n D^*D^*} 
\psi_n^{\mu}\bigg \{  
-D_i^{*+\alpha}\!\stackrel{\leftrightarrow}{\partial}_{\mu}D^{*}_{i\alpha} 
+D_{i\mu}^{*+}\!\stackrel{\leftrightarrow}{\partial}_{\alpha}D^{*\alpha}_{i} 
+D_i^{*+\alpha}\!\stackrel{\leftrightarrow}{\partial}_{\alpha}D^{*}_{i\mu}
\bigg \} 
\nonumber\,, 
\en 
where $\psi_n$ with $n=1,2$ stands for the vector fields of $J/\psi$ and  
$\psi(2S)$, respectively. The chiral couplings of $\psi_n$ with $D(D^*)$  
mesons read  
\begin{eqnarray} 
g_{\psi_n DD}=g_{\psi_n D^*D^*} \frac{m_D}{m_{D^*}} =  
g_{\psi_n DD^*} m_{\psi_n} \sqrt{\frac{m_D}{m_{D^*}}} =  
\frac{m_{\psi_n}}{f_{\psi_n}} \; .  
\en 
The quantity $f_{\psi_n}$ is determined by the leptonic decay  
widths of $J/\psi$ and $\psi(2S)$ of  
\begin{eqnarray}\label{epem_widths} 
\Gamma(J/\psi\rightarrow e^+e^-) 
&=&\frac{16\pi}{27}\frac{\alpha^2}{m_{J/\psi}}
f_{J/\psi}^2=5.55 \ {\rm keV} \; ,\nonumber \\ 
\Gamma(\psi(2S)\rightarrow e^+e^-) 
&=&\frac{16\pi}{27}\frac{\alpha^2}{m_{\psi(2S)}} 
f_{\psi(2S)}^2=2.38 \ {\rm keV} \,,  
\end{eqnarray} 
where $\alpha$ is the fine structure constant. 
From Eq.~(\ref{epem_widths}) we get 
$f_{J/\psi} = 416.4$ MeV and $f_{\psi(2S)} = 297.5$ MeV. 

Thus the ratios of the coupling constants $g_{\psi(2S)DD}$  
and $g_{\psi(2S)D^*D^*}$ to $g_{J/\psi DD}$ and $g_{J/\psi D^*D^*}$ are
fixed as
\begin{eqnarray} 
\frac{g_{\psi(2S) DD}}{g_{J/\psi DD}} =  
\frac{g_{\psi(2S) D^*D^*}}{g_{J/\psi D^*D^*}} =  
\frac{m_{\psi(2S)}}{m_{J/\psi}} \, \frac{f_{J/\psi}}{f_{\psi(2S)}}  
\simeq 1.67 \,. 
\end{eqnarray} 
The coupling constant $g_{D^*D\gamma}$ is fixed by data (central  
values) on the radiative decay widths  
$\Gamma(D^* \to D\gamma)$~\cite{Nakamura:2010zz} with:  
\eq\label{gamma_rad}  
\Gamma(D^{\ast \, +} \to D^+ \gamma) &=& \frac{\alpha}{24} \,  
g_{_{D^{\ast \, -}D^+\gamma}}^2 \, m_{D^{\ast \, +}}^3 \,  
\biggl( 1 - \frac{m_{D^{+}}^2}{m_{D^{\ast \, +}}^2} \biggr)^3 
= 1.54 \, {\rm keV } \,, \nonumber\\  
\Gamma(D^{\ast \, 0} \to D^0 \gamma) &=& \frac{\alpha}{24} \,  
g_{_{D^{\ast \, 0}D^0\gamma}}^2 \, m_{D^{\ast \, 0}}^3 \,  
\biggl( 1 - \frac{m_{D^{0}}^2}{m_{D^{\ast \, 0}}^2} \biggr)^3 
= 26.04 \ {\rm keV } \,.  
\en  
From Eq.~(\ref{gamma_rad}) we finally predict  
\eq  
g_{D^{*-}D^+\gamma} = 0.5 \ {\rm GeV}^{-1}\,, \hspace*{.5cm}   
g_{D^{*0}D^0\gamma} = 2.0 \ {\rm GeV}^{-1}\,. \hspace*{.5cm}   
\en  
In our calculation the mass of the charm quark is chosen as
$m_c\simeq m_X/2$. The respective couplings $g_{V\gamma}$ ($V=\rho^0,\omega$)
of the transitions $\rho\gamma$ and $\omega\gamma$
are fixed from
data on the $V \to e^+e^-$ decay widths as    
\eq\label{GammaVee}  
\Gamma(\rho^0 \to e^+e^-)  &=& \frac{4\pi}{3} \alpha^2 g_{\rho\gamma}^2 m_\rho 
= 7.04 \ {\rm keV} \,, \nonumber\\  
\Gamma(\omega \to e^+e^-)  &=& \frac{4\pi}{3} \alpha^2  
g_{\omega\gamma}^2 m_\omega = 0.60 \ {\rm keV} \,.  
\en  
Using last equation we get $g_{\rho\gamma} = 0.20$ and  
$g_{\omega\gamma} = 0.06$.   
Finally, the interaction Lagrangian for the coupling of $\psi_n$ to the
$c\bar{c}$ configuration is 
\begin{eqnarray} 
{\cal L}_{\psi_nc\bar{c}}(x)=g_{\psi_n}\psi^{\mu}_n(x) 
\int d^4y \Phi_{\psi_n}(y^2)
\bar{c}(x+y/2)\gamma_{\mu}c(x-y/2), 
\end{eqnarray} 
where the coupling constant $g_{\psi_n}$ is given as 
\begin{eqnarray}  
g_{\psi_n}=\frac23\frac{m_{\psi_n}}{f_{\psi_n}}. 
\end{eqnarray}    
By using the decay constants $f_{\psi_n}$ of the $J/\psi$ and $\psi(2S)$  
states we obtain the values $g_{J/\psi} = 4.96$ and $g_{\psi(2S)} = 8.26$. 
Here $\Phi_{\psi_nc}$ is the vertex function corresponding to the distribution 
of the charm quark inside the $\psi_n$. 
For the $J/\psi$ and its radially excited 
state $\psi(2S)$ we use the respective pure and modified Gaussians 
as correlation functions. Their Fourier transforms in the 
Euclidean region read:  
\eq 
\tilde\Phi_{J_\psi}(p_E^2/\Lambda^2_{J_\psi}) &\doteq& 
\exp( - p_E^2/\Lambda_{J_\psi}^2)\,, 
\nonumber\\
\tilde\Phi_{\psi}(p_E^2/\Lambda^2_I) &\doteq& 
\exp( - p_E^2/\Lambda_{\psi}^2) (1 + \alpha_{\psi} p_E^2/\Lambda_{\psi}^2) 
\,, 
\en 
where $\Lambda_{J_\psi} = \Lambda_{\psi} = 2.5$ GeV is a dimensional 
parameter describing the distribution of charm quarks in the corresponding 
charmonia states, and $\alpha_\psi$ is a free parameter, 
which is fixed by matching the present results for the radiative decays to the
ones of $c\bar c$ potential models.
The choice for the dimensional parameter $\Lambda_{J_\psi, ~\psi}$ is 
not unique, an enhanced value of for example 3 GeV would result in 
a rescaled quantity $\alpha_\psi$, but the final result remains unchanged.

\section{Results}  
 
In the following we define the
the binding energy $\epsilon=\epsilon_{D^0D^{*0}}$ by setting   
\begin{eqnarray} 
m_X=M_{D^0}+M_{D^{0*}}-\epsilon_{D^0D^{*0}}. 
\end{eqnarray} 
We present our numerical results for three typical values of  
this binding energy $\epsilon = 0.3, 0.7, 1$ MeV  
and use the corresponding sets of configuration probabilities  
as calculated in Ref.~\cite{Swanson:2003tb}:  
\eq\label{ZH1H2_factors_03} 
Z_{\d\ds} = 0.92\,, \hspace*{.25cm}  
Z_{\dppm\dpmps} = 0.033\,, \hspace*{.25cm}  
Z_{\jpsi\omega} = 0.041\,, \hspace*{.25cm}  
Z_{\jpsi\rho} = 0.006
\en   
for $\epsilon = 0.3$ MeV,  
\eq\label{ZH1H2_factors_07}  
Z_{\d\ds} = 0.816\,, \hspace*{.25cm}  
Z_{\dppm\dpmps} = 0.079\,, \hspace*{.25cm}  
Z_{\jpsi\omega} = 0.096\,, \hspace*{.25cm}  
Z_{\jpsi\rho} = 0.009
\en   
for $\epsilon = 0.7$ MeV, and  
\eq\label{ZH1H2_factors_10}  
Z_{\d\ds} = 0.758\,, \hspace*{.25cm}  
Z_{\dppm\dpmps} = 0.111\,, \hspace*{.25cm}  
Z_{\jpsi\omega} = 0.122\,, \hspace*{.25cm}  
Z_{\jpsi\rho} = 0.009
\en   
for $\epsilon = 1$ MeV.  

For the explicit discussion of our results
we display the leading contributions
of the different structure components
(molecular or charmonium) to the
$X\rightarrow J/\psi+\gamma$ and $X\rightarrow\psi(2S)+\gamma$ decay widths
in powers of $P_{J_\psi(\psi)}/m_X$ with
$m_X \sim m_{J_\psi} \sim m_{\psi(2S)}$
(in the further discussion we will focus on the
case of $\epsilon = 0.3$ MeV; the full analysis given in Tables I--III
includes the other cases of
$\epsilon = 0.7$ and 1 MeV):
\eq\label{ratio_leading}
\hspace*{-2cm}
\Gamma(X \to J/\psi+\gamma) &=&
\frac{\alpha P_{J_\psi}^3}{3m_X^2} \ \Biggl[\biggl(G^{DD^*}_{J_\psi}
\, \cos\theta \ \frac{P_{J_\psi}}{m_X}
+ G^{J_\psi V} \, \cos\theta
+ G^{c\bar c}_{J_\psi} \, \sin\theta  \biggr)^2 \nonumber\\
&+& \biggl(G^{J_\psi V} \, \cos\theta 
+G^{c\bar c}_{\psi} \, \sin\theta \biggr)^2\frac{m_X^2}{m_{J_\psi}^2} \Biggr]
\,, \\
\hspace*{-2cm}
\Gamma(X \to \psi(2s)+\gamma) &=&
\frac{\alpha P_{\psi}^3}{3 m_X^2} \
\Biggl[\biggl(G^{DD^*}_{\psi} \, \cos\theta  \ \frac{P_{\psi}}{m_X}
+ G^{c\bar c}_{\psi} \, \sin\theta \biggr)^2
+ \biggl(G^{c\bar c}_{\psi} \, \sin\theta\biggr)^2 \frac{m_X^2}{m_{J_\psi}^2}
\Biggr] \,,\nonumber 
\en
where $P_{J_\psi} = (m_X^2 -  m_{J_\psi}^2)/(2 m_X) = 697$~MeV
and $P_{\psi} = (m_X^2 -  m_{\psi}^2)/(2 m_X) = 181$~MeV
are the corresponding three--momenta of the decay products;
$\theta$ is the mixing angle between the hadronic and the charmonium
components. Here the coefficients $G_i^j$ are the effective dimensionless
couplings containing the leading contributions of the different
$X(3872)$ subconfigurations to the radiative decay widths. In particular,
$G^{DD^*}_{J_\psi(\psi)}$ are the couplings
encoding the contributions of the molecular
$D^0D^{\ast 0} + D^\pm D^{\ast\mp}$ components,
$G^{J_\psi V}$ is related to
the $J/\psi \rho$ and $J/\psi \omega$ components
and 
$G^{c \bar c}_{J_\psi(\psi)}$ are the couplings
connected with the possible charmonium configuration in the $X(3872)$.
Note for convenience we define the relative signs of the different couplings
in the Lagrangian~(\ref{Lagr_X}) in a such way, that all contributions of the
$\d\ds$, $J/\psi V$ and $c\bar c$ configurations enter with positive
relative signs. However, we want to stress that the signs of the relative
couplings are not fixed at the beginning and they can at this level
of the analysis only be fixed by an analysis of the radiative decay data.

First, we consider the radiative decays of the $X(3872)$ in the pure 
charmonium picture. To fix the model parameters, entering in the charmonia 
vertex functions, we match our results to the model predictions performed in
Refs.~\cite{Barnes:2003vb,Barnes:2005pb,Li:2009zu}. 
Since these predictions cover a wider range of values we explicitly refer
to the results of~\cite{Barnes:2003vb}, \cite{Barnes:2005pb} 
and \cite{Li:2009zu} as scenario I, II and III, respectively. 
The predictions of these scenarios for the rates 
$\Gamma_{\psi} = \Gamma(X \to \psi(2S)+\gamma)$ and 
$\Gamma_{J_\psi} = \Gamma(X \to J/\psi+\gamma)$ and 
their ratio $R = \Gamma_{\psi}/\Gamma_{J_\psi}$ are contained in Table I. 
In order to reproduce (or match) the predictions of the potential approaches 
we adjust the two parameters $\alpha_X$ and $\alpha_\psi$ entering 
in the correlation 
functions of the charmonium components for the $X(3872)$ and the $\psi(2S)$. 
In Table I we present the sets of parameters 
$\alpha_X$ and $\alpha_\psi$ fixed from this matching procedure.
We also show the sensitivity of the parameter values to a variation of the 
binding energy $\epsilon$.

The inclusion of the molecular 
components of the $X(3872)$ can be relevant to improve first the
description of the ratio $R$. 
However, due to the large error attached to the experimental result 
for the ratio $R$ and due to the band width of predictions for the pure 
charmonium case
we only can perform a preliminary structure analysis of the $X(3872)$. 
Restricting to the current central value of the ratio with $R=3.5$ 
we demonstrate that an inclusion of the molecular components 
can decrease (as is the case for scenario I) or increase (as for
scenarios II and III) the potential results of the pure $c\bar c$ 
configurations. In particular, 
for case I the contribution of 
both molecular components $\d\ds$ and $J/\psi V$ to the $J/\psi \gamma$ 
decay mode should be constructive 
relative to the charmonium one in order to suppress the 
ratio $R$. For the cases II and III the molecular 
components $\d\ds$ and $J/\psi V$ should result in destructive interference 
with the charmonium component in order 
to enlarge the ratio $R$. It should also be clear that the 
constructive/destructive interference effect can be controlled by 
the sign of the mixing angle $\theta$. 

In Tables II-IV we present a detailed analysis of
the radiative decay widths and the resulting ratio $R$
for the full structure scenario of the $X(3872)$ containing
both molecular configurations and a charmonium component.
For the case of the $c\bar c$ configuration all three scenarios
contained in Table I are considered. We also display the various
cases for the binding energy with $\epsilon = 0.3, 0.7$ and 1~MeV,
connected with the varying molecular configuration probabilities
of Eqs.~(\ref{ZH1H2_factors_03})--(\ref{ZH1H2_factors_10}).
Note again that the charmonium contribution is fitted 
by an appropriate choice of the parameters $\alpha_X$ and $\alpha_\psi$ 
in the vertex functions of the $X(3872)$ and $\psi(2S)$ states describing 
the distribution of the charm quark components inside these states. 
We also indicate the value of the mixing angle $\theta$ to reproduce
the central value $R = 3.5$ for all three scenarios and for the representative
values of the binding energy $\epsilon$.
As was already stressed, for scenario I inclusion of the molecular 
components should induce a decrease of the ratio $R$ predicted in 
the pure charmonium 
picture. The interference of the molecular and charmonium components 
for the $J/\psi \gamma$ decay amplitude is constructive
and the corresponding mixing angle $\theta$ is positive and 
relatively large with $\theta \sim 70^0$ ($\sin\theta \simeq 0.94$). 
Therefore, in this scenario a possible molecular contribution is largely 
suppressed. In the case of scenarios II and III inclusion 
of the molecular components leads to an increase of the value $R$
taken in the pure charmonium picture. The interference of the molecular and 
charmonium components in the $J/\psi \gamma$ decay is destructive and 
the corresponding mixing angle is small and negative with 
$\theta \simeq -13^0$ ($\sin\theta \simeq - 0.23$, scenario II) and 
$\theta \simeq -20^0$ ($\sin\theta \simeq -0.34$, scenario III). 
Finally, we would like to comment on a possible deviation of 
the cutoff parameters for the neutral $D^0 \bar D^{\ast 0}$ 
and charged $D^\pm \bar D^{\ast \mp}$ components. 
Because the charged component is strongly suppressed even a large variation
in the cutoff parameters does not have a strong influence on the result.
For example, using a cutoff parameter for the $D^\pm \bar D^{\ast \mp}$ component 
from 0.5 to 2 GeV we find that the total contribution of the neutral 
$D^0 \bar D^{\ast 0}$ and charged $D^\pm \bar D^{\ast \mp}$ components to 
the radiative decay widths of the X(3872) are varied respectively:
$\Gamma_{J_\psi}$ from 3.6 to 4.5 keV and
$\Gamma_\psi$ from 0.012 to 0.015 keV for $\epsilon = 0.3$ MeV. Similar 
results are obtained for $\epsilon = 0.7$ and $1$ MeV.

Next we discuss the predictions of our approach for the strong decays 
of the $X(3872)$. In particular, we consider the transitions 
$X(3872) \to J/\psi + (2\pi, 3\pi)$ and 
$X(3872) \to \chi_{cJ} + (\pi, 2\pi)$ where a preliminary evaluation and
discussion was given in Ref.~\cite{Dong:2009gb}. There numerical results 
were indicated for large
values of the cutoff parameters for the molecular components in the $X(3872)$.
Now we present predictions for these strong decay modes with values
for the model parameters as fixed in the present manuscript. 
As in Ref.~\cite{Dong:2009gb} we restrict to the contribution of the 
molecular $J/\psi\rho$ and $J/\psi\omega$ components to the 
transitions $X(3872) \to J/\psi + (2\pi, 3\pi)$ (see diagrams in Fig.3),
while the $D^0 \bar D^{\ast 0}$ and charged $D^\pm \bar D^{\ast \mp}$ components
contribute to the
$X(3872) \to \chi_{cJ} + (\pi, 2\pi)$ transitions (see diagrams 
in Figs.4 and 5). 
Especially the strong decay modes involving $J/\psi$ in the final state
are interesting, since the Belle~\cite{Abe:2005ix}
and {\it BABAR}~\cite{Aubert:2008rn} Collaborations provided 
experimental results for the following ratios:
\eq\label{data_strong}
\hspace*{-2cm}
R_1 = \displaystyle\frac{\Gamma(X \to \jp \pi^+ \pi^- \pi^0)}
{\Gamma(X \to \jp \pi^+ \pi^-)}= 1.0 \pm 0.4 \ ({\rm stat})   
\pm 0.3 \ ({\rm syst})   
\en 
and 
\eq\label{data_radiative}
\hspace*{-2cm}
R_2 = \displaystyle\frac{\Gamma(X \to \jp\gamma)}
{\Gamma(X \to \jp\pi^+\pi^-)}= 0.14 \pm 0.05 \ \ ({\rm Belle}); \ 
                             0.33 \pm 0.12 \ \ ({\it BABAR}) \,. 
\en  
In Tables V we present our results for the strong decay modes of the $X(3872)$,
where we also indicate the values obtained for $R_1$ and by $R_{cJ}$ 
we define the ratios of three-- and two--body decays with 
\eq 
R_{cJ} = 
\displaystyle\frac{\Gamma(X \to \chi_{cJ} + 2\pi^0)}
{\Gamma(X \to \chi_{cJ} + \pi^0)} \,. 
\en  
Table VI contains our predictions for the 
$X(3872) \to J/\psi \gamma$ decay, but now also for the ratio $R_2$ 
relating strong and radiative decays. 
The ratio $R_2$ can be explained in the model scenarios II and III, 
where only a subleading $c\bar c$ component in the $X(3872)$  is present.
One should stress that all predictions are sensitive to the $Z$-factors 
($Z_{\d\ds}$, $Z_{\dppm\dpmps}$, $Z_{\jpsi\omega}$ and $Z_{\jpsi\rho}$) 
encoding the relevant contribution of the different components in
the $X(3872)$ Fock state. As was already stated, in our numerical 
calculations we use the explicit predictions of the potential 
model~\cite {Swanson:2003tb}.

\section{Summary}  
 
In summary, we have shown that a nontrivial interplay between
a possible charmonium and the molecular components in the $X(3872)$
is necessary to explain the ratio
$R = {\cal B}(X(3872)\rightarrow \psi(2S)\gamma)/
{\cal B}(X(3872)\rightarrow J/\psi\gamma)$ of radiative decay modes.
Since present experimental result has a fairly large error and
the $c\bar c$ predictions are also sensitive to model details,
at this point the structure composition of the $X(3872)$ related to this ratio
cannot be pinned down uniquely.
But we also showed that especially the ratio 
$R_2 = {\Gamma(X \to \jp\gamma)}/
{\Gamma(X \to \jp\pi^+\pi^-)}$ relating radiative to strong decays 
hints presently
at a scenario where the $c\bar c$ component only plays a subleading role.
More precise data on these ratios obviously help in further constraining
the structure details of the $X(3872)$.
Further tests of the proposed interpretation of the $X(3872)$ can also
be found in the predictions for the strong decay modes involving $\chi_{cJ}$.  

\section*{Acknowledgments}

This work was supported by the DFG under Contract
No. FA67/31-2 and No. GRK683. 
This work is supported  by the National Sciences Foundations  
No. 10775148 and 10975146 and by CAS grant No. KJCX3-SYW-N2 (YBD).  
This research is also part 
of the European Community-Research Infrastructure Integrating Activity 
``Study of Strongly Interacting Matter'' (HadronPhysics2, Grant Agreement 
No. 227431), Russian President grant ``Scientific Schools'' No. 3400.2010.2, 
Federal Targeted Program "Scientific and scientific-pedagogical personnel 
of innovative Russia" Contract No. 02.740.11.0238. 
Y.B.D. would like to thank the T\"ubingen theory group for its hospitality.   
V.E.L. would like to thank the Theory Group of the Institute of High Energy   
Physics (Beijing) for its hospitality.

\newpage  

\begin{figure}[htb]  

\begin{center} 
\epsfig{file=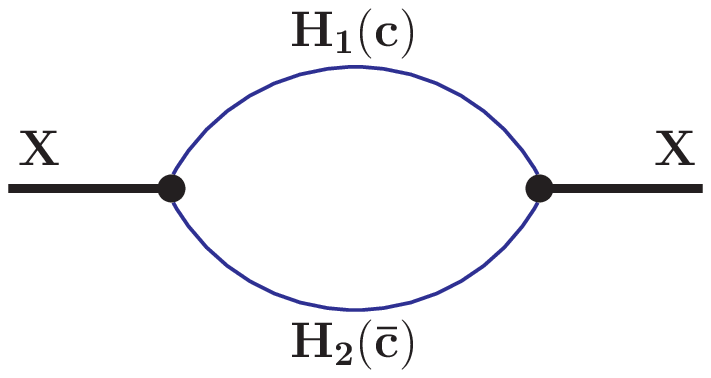, scale=.7} 
\end{center} 
\caption{$H_1H_2$ hadron--loop or $c \bar c$ quark--loop diagram  
contributing to the mass operator of the $X(3872)$ meson.}

\begin{center} 
\epsfig{file=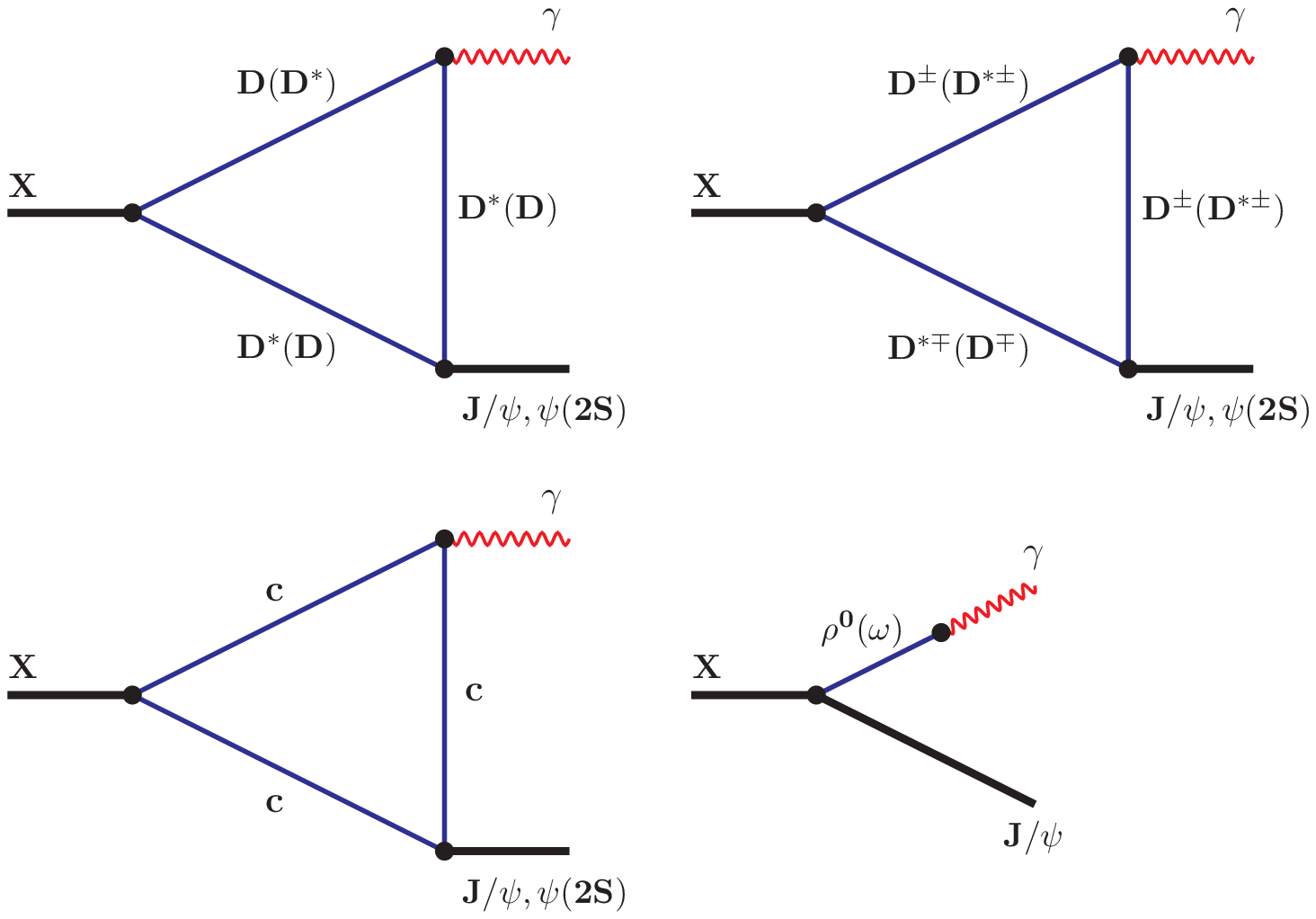, scale=.7} 
\end{center} 
\caption{Diagrams contributing to the radiative transitions  
$X(3872) \to J/\psi + \gamma$ and $X(3872) \to \psi(2S) + \gamma$.} 

\begin{center}
\epsfig{file=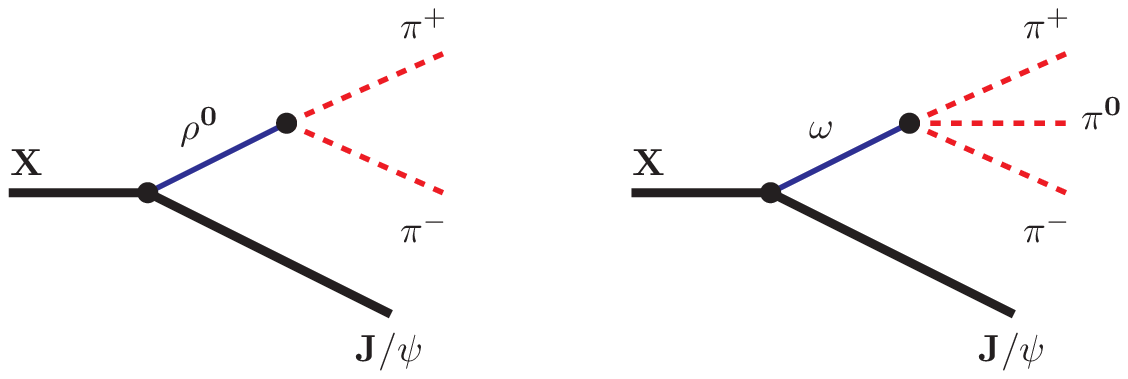, scale=.7}
\end{center}
\caption{Diagrams contributing to the hadronic transitions
$X(3872) \to J/\psi + (2\pi, 3\pi)$.} 

\end{figure}

\newpage 

\begin{figure}[htb] 
\begin{center}
\epsfig{file=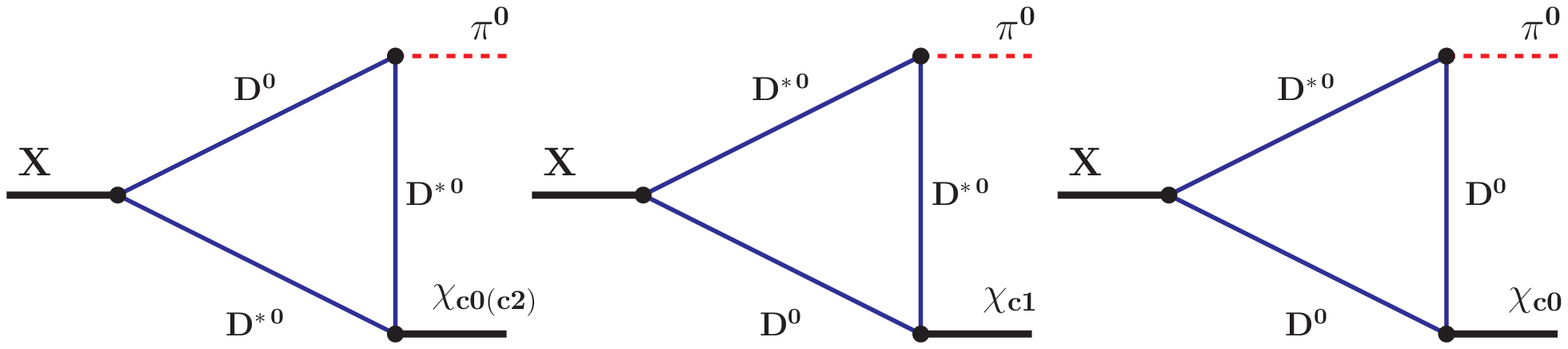, scale=.7}
\end{center}
\caption{Diagrams contributing to the hadronic transitions
$X(3872) \to \chi_{cJ} + \pi^0$.}
%\end{figure}

%\newpage

%\begin{figure}[htb] 
\begin{center}
\epsfig{file=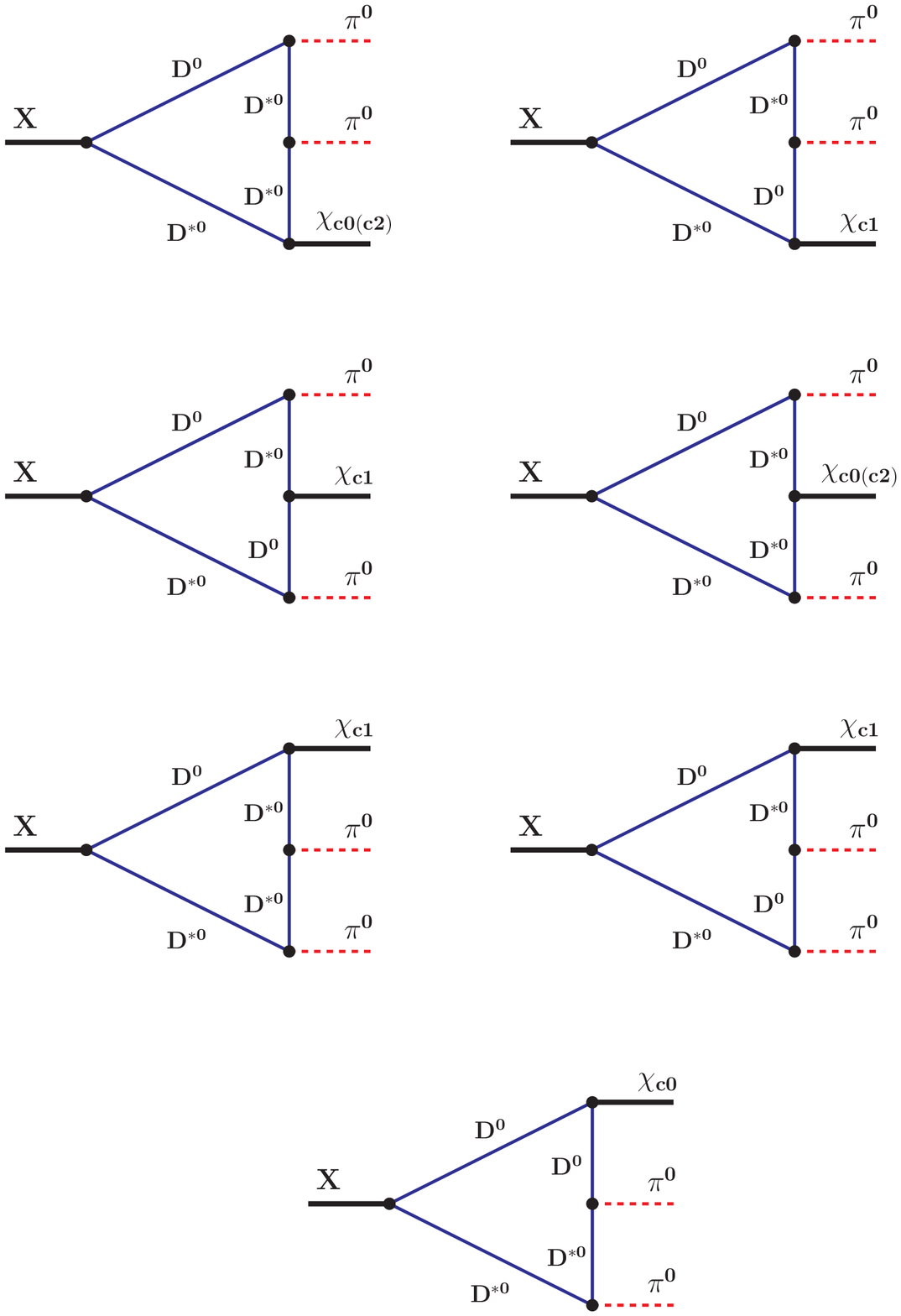, scale=.7}
\end{center}
\caption{Diagrams contributing to the hadronic transitions
$X(3872) \to \chi_{cJ} + 2 \pi^0$.}
\end{figure}

\newpage 

\begin{center}
{\bf Table I.}
Radiative decay widths of $X(3872)$ for the pure $2^3P_1$ $c\bar c$ case.\\
The matching parameters $\alpha_X$ and $\alpha_\psi$ are indicated resulting  
from a fit to \\ the decay widths of the respective potential models
(scenarios I, II and III).

\vspace*{.25cm}
\def\arraystretch{1.5}
\begin{tabular}{|c|l|l|l|} \hline
&   Scenario I~\cite{Barnes:2003vb}  
&   Scenario II~\cite{Barnes:2005pb} 
&   Scenario III~\cite{Li:2009zu} \\
\cline{2-4}
    & \ $\Gamma_{\psi}   = 64$  keV
    & \ $\Gamma_{\psi}   = 183$ keV
    & \ $\Gamma_{\psi}   = 60$  keV \\
%\cline{2-4}
    & \ $\Gamma_{J_\psi} = 11$  keV \ 
    & \ $\Gamma_{J_\psi} = 71$  keV 
    & \ $\Gamma_{J_\psi} = 45$  keV \\
%\cline{2-4}
    & \ \ \ \ \ $R = 5.8$ 
    & \ \ \ \ \ $R = 2.6$ 
    & \ \ \ \ \ $R = 1.3$ \\
\hline\hline
  $\epsilon$ (in MeV) 
& \ \ \ \ \ $\alpha_X$, $\alpha_\psi$ 
& \ \ \ \ \ $\alpha_X$, $\alpha_\psi$ 
& \ \ \ \ \ $\alpha_X$, $\alpha_\psi$ \\
\cline{1-4} 
0.3 & \ \ 2.094, -2.390 & \ \ 1.685, -2.750 & \ \ 1.814, -2.114 \\ 
\cline{1-4} 
0.7 & \ \ 2.094, -2.394 & \ \ 1.684, -2.757 & \ \ 1.813, -2.118 \\ 
\cline{1-4} 
1   & \ \ 2.094, -2.398 & \ \ 1.684, -2.762 & \ \ 1.813, -2.122 \\
\cline{1-4} 
\end{tabular}
\end{center}

\vspace*{.5cm}
 
\begin{center}
{\bf Table II.}
Radiative decay widths of $X(3872)$ in the charmonium-molecular picture\\ 
for a binding energy of $\epsilon = 0.3$~MeV. The total result includes \\ 
all contributions with the appropriate mixing factors 
$\cos\theta$ or $\sin\theta$ of Eq.(\ref{Xstate}). \\ 
The value of $\theta$ is indicated in the bracket after the total prediction 
for $R$. 

\vspace*{.25cm}
\def\arraystretch{1.5}
\begin{tabular}{|c|c|c|c|c|c|c|} \hline
  Scenario 
& Quantity 
& $c\bar c$ 
& $D D^\ast$ 
& $J/\psi V$ 
& $D D^\ast + J/\psi V$ 
& Total \\ 
\hline\hline  
& $\Gamma_{J_\psi}$, keV  
& 11
& 3.6 
& 1.5
& 8  
& 16
\\ 
\cline{2-7} 
  I
& $\Gamma_\psi$, keV   
& 64
& 0.01
& 0
& 0.01
& 56 
\\ 
\cline{2-7}   
& $R$ 
& 5.8 
& 3.3 $\times 10^{-3}$  
& 0 
& 1.5 $\times 10^{-3}$  
& 3.5 ($\theta = 68.9^0$) 
\\ \hline\hline  
& $\Gamma_{J_\psi}$, keV  
& 71
& 3.6 
& 1.5
& 8  
& 2.4
\\ 
\cline{2-7} 
  II
& $\Gamma_\psi$, keV   
& 183
& 0.01
& 0
& 0.01
& 8.4 
\\ 
\cline{2-7}   
& $R$ 
& 2.6 
& 3.3 $\times 10^{-3}$  
& 0 
& 1.5 $\times 10^{-3}$  
& 3.5 ($\theta = - 12.6^0$) 
\\ \hline\hline  
& $\Gamma_{J_\psi}$, keV  
& 45
& 3.6 
& 1.5
& 8  
& 1.94
\\ 
\cline{2-7} 
  III 
& $\Gamma_\psi$, keV   
& 64
& 0.01
& 0
& 0.01
& 6.8
\\ 
\cline{2-7}   
& $R$ 
& 5.8 
& 3.3 $\times 10^{-3}$  
& 0 
& 1.5 $\times 10^{-3}$  
& 3.5 ($\theta = -20.2^0$) 
\\ 
\hline 
\hline 
\end{tabular}
\end{center}

\vspace*{1cm} 

\newpage
 
\begin{center}
{\bf Table III.}
Radiative decay widths of $X(3872)$ in the charmonium-molecular picture\\
for a binding energy of $\epsilon = 0.7$~MeV. Otherwise as in Table II.

\vspace*{.25cm}
\def\arraystretch{1.5}
\begin{tabular}{|c|c|c|c|c|c|c|} \hline
  Scenario 
& Quantity 
& $c\bar c$ 
& $D D^\ast$ 
& $J/\psi V$ 
& $D D^\ast + J/\psi V$ 
& Total \\ 
\hline\hline  
& $\Gamma_{J_\psi}$, keV  
& 11
& 3.4
& 3.1
& 10.6  
& 16.6
\\ 
\cline{2-7} 
  I
& $\Gamma_\psi$, keV   
& 64
& 0.01
& 0
& 0.01
& 58.2 
\\ 
\cline{2-7}   
& $R$ 
& 5.8 
& 3.2 $\times 10^{-3}$  
& 0 
& 1.0 $\times 10^{-3}$  
& 3.5 ($\theta = 72.1^0$) 
\\ \hline\hline  
& $\Gamma_{J_\psi}$, keV  
& 71
& 3.4 
& 3.1
& 10.6 
& 2.7
\\ 
\cline{2-7} 
  II
& $\Gamma_\psi$, keV   
& 183
& 0.01
& 0
& 0.01
& 9.5
\\ 
\cline{2-7}   
& $R$ 
& 2.6 
& 3.2 $\times 10^{-3}$  
& 0 
& 1.0 $\times 10^{-3}$  
& 3.5 ($\theta = - 13.5^0$) 
\\ \hline\hline  
& $\Gamma_{J_\psi}$, keV  
& 45
& 3.4 
& 3.1
& 10.6  
& 2.0
\\ 
\cline{2-7} 
  III 
& $\Gamma_\psi$, keV   
& 60
& 0.01
& 0
& 0.01
& 7.0
\\ 
\cline{2-7}   
& $R$ 
& 5.8 
& 3.2 $\times 10^{-3}$  
& 0 
& 1.0 $\times 10^{-3}$  
& 3.5 ($\theta = -20.4^0$) 
\\ 
\hline 
\hline 
\end{tabular}
\end{center}

\vspace*{.5cm}
 
\begin{center}
{\bf Table IV.}
Radiative decay widths of $X(3872)$ in the charmonium-molecular picture \\
for a binding energy of $\epsilon = 1.0$~MeV. Otherwise as in Table II.

\vspace*{.25cm}
\def\arraystretch{1.5}
\begin{tabular}{|c|c|c|c|c|c|c|} \hline
  Scenario 
& Quantity 
& $c\bar c$ 
& $D D^\ast$ 
& $J/\psi V$ 
& $D D^\ast + J/\psi V$ 
& Total \\ 
\hline\hline  
& $\Gamma_{J_\psi}$, keV  
& 11
& 3.3 
& 3.7
& 11.4  
& 16.8
\\ 
\cline{2-7} 
  I
& $\Gamma_\psi$, keV   
& 64
& 0.01
& 0
& 0.01
& 58.8 
\\ 
\cline{2-7}   
& $R$ 
& 5.8 
& 3.2 $\times 10^{-3}$  
& 0 
& 0.9 $\times 10^{-3}$  
& 3.5 ($\theta = 72.9^0$) 
\\ \hline\hline  
& $\Gamma_{J_\psi}$, keV  
& 71
& 3.3 
& 3.7
& 11.4
& 2.8
\\ 
\cline{2-7} 
  II
& $\Gamma_\psi$, keV   
& 183
& 0.01
& 0
& 0.01
& 9.8 
\\ 
\cline{2-7}   
& $R$ 
& 2.6 
& 3.2 $\times 10^{-3}$  
& 0 
& 0.9 $\times 10^{-3}$  
& 3.5 ($\theta = - 13.8^0$) 
\\ \hline\hline  
& $\Gamma_{J_\psi}$, keV  
& 45
& 3.3 
& 3.7
& 11.4  
& 2.0
\\ 
\cline{2-7} 
  III 
& $\Gamma_\psi$, keV   
& 64
& 0.01
& 0
& 0.01
& 7.0
\\ 
\cline{2-7}   
& $R$ 
& 5.8 
& 3.2 $\times 10^{-3}$  
& 0 
& 0.9 $\times 10^{-3}$  
& 3.5 ($\theta = -20.6^0$) 
\\ 
\hline 
\hline 
\end{tabular}
\end{center}

\newpage 

\begin{center}
{\bf Table V.} 
Predictions for the strong decays of $X(3872)$. 

\vspace*{.25cm}

\hspace*{-.5cm}
\def\arraystretch{1.4}
\begin{tabular}{|c|c|c|c|c|}  
\hline
Quantity & \multicolumn{3}{c|}{Our results} & Data \\
           \cline{2-4}
         & $\epsilon = 0.3$ MeV 
         & $\epsilon = 0.7$ MeV 
         & $\epsilon = 1$ MeV 
         & \\
\hline 
$\Gamma(X \to \jp \pi^+ \pi^-)$, keV & 6.0 & 9.7 & 10.5 &\\ 
\hline
$\Gamma(X \to \jp \pi^+ \pi^- \pi^0)$, keV & 3.6 & 7.4 & 11.2 &\\ 
\hline
$R_1$ & 0.60 & 0.76 & 1.07 & $1.0 \pm 0.4 \pm 0.3$ \\ 
\hline 
$\Gamma(X \to \chi_{c0} + \pi^0)$, keV &2.76 &3.17 &3.38 & \\ 
\hline 
$\Gamma(X \to \chi_{c0} + 2\pi^0)$, eV &4.24 &4.88 &5.21 & \\ 
\hline 
$R_{c0} \times 10^3$ &1.54 &1.54 &1.54 & \\ 
\hline 
$\Gamma(X \to \chi_{c1} + \pi^0)$, keV &0.74 &0.86 &0.92 & \\ 
\hline
$\Gamma(X \to \chi_{c1} + 2\pi^0)$, eV &49.84 &57.34 &61.17 & \\ 
\hline 
$R_{c1} \times 10^2$ &6.70 &6.70 &6.70 & \\ 
\hline 
$\Gamma(X \to \chi_{c2} + \pi^0)$, keV &1.01 &1.16 &1.23 & \\ 
\hline
$\Gamma(X \to \chi_{c2} + 2\pi^0)$, eV &1.39 &1.59 &1.70 & \\
\hline 
$R_{c2} \times 10^3$ &1.38 &1.37 &1.38 & \\ 
\hline 
\end{tabular}
\end{center}

\vspace*{.5cm} 

\begin{center}
{\bf Table VI.} 
Results for $X(3872) \to J/\psi \gamma$ and the ratio $R_2$. 

\vspace*{.25cm}

\def\arraystretch{1.4}
\begin{tabular}{|c|c|c|c|c|c|}  
\hline
Scenario & Quantity & \multicolumn{3}{c|}{Our results} & Data \\
           \cline{3-5}
         &
         & $\epsilon = 0.3$ MeV 
         & $\epsilon = 0.7$ MeV 
         & $\epsilon = 1$ MeV 
         & \\
\hline 
 & $\Gamma_{J_\psi}$, keV & 16 & 16.6 & 16.8 & \\ 
\cline{2-6}
I  & $R_2$ & 2.67 & 1.71 & 1.57 & 0.14 $\pm$ 0.05~\cite{Abe:2005ix} \\ 
  &   & & & & 0.33 $\pm$ 0.12~\cite{Aubert:2008rn}\\
\hline 
 & $\Gamma_{J_\psi}$, keV & 2.4 & 2.7 & 2.8 & \\ 
\cline{2-6}
II   & $R_2$ & 0.67 & 0.36 & 0.25 & 0.14 $\pm$ 0.05~\cite{Abe:2005ix} \\ 
   &  & & & & 0.33 $\pm$ 0.12~\cite{Aubert:2008rn}\\
\hline 
 & $\Gamma_{J_\psi}$, keV & 1.9 & 2.0 & 2.0 & \\ 
\cline{2-6}
III    & $R_2$ & 0.53 & 0.27 & 0.18 & 0.14 $\pm$ 0.05~\cite{Abe:2005ix} \\ 
   & & & & & 0.33 $\pm$ 0.12~\cite{Aubert:2008rn}\\
\hline 
\end{tabular}
\end{center}

\end{document}